\documentstyle[art12,epsf,makeidx,bezier]{article}
\newcommand{\lb}[1]{\label{eqn:#1}}
\newcommand{\rf}[1]{\ref{eqn:#1}}
\begin{document}
\baselineskip 20pt
\vglue 2cm
\centerline{\Large {\bf A Symmetric Generalization}}
\centerline{\Large {\bf of Linear B\"acklund Transformation}}
\centerline{\Large {\bf associated with the Hirota Bilinear Difference
Equation}}

\vglue 4cm
\centerline{Nobuhiko SHINZAWA \quad and \quad Satoru SAITO}
\vglue 1cm
\centerline{{\it Department of Physics, Tokyo Metropolitan University}}
\centerline{{\it Minamiohsawa 1-1, Hachiohji, Tokyo 192-03, Japan}}
\vglue 6cm
\begin{abstract}
The Hirota bilinear difference equation is generalized to discrete space of
arbitrary dimension. Solutions to the nonlinear difference equations can be
obtained via B\"acklund transformation of the corresponding linear problems.
\end{abstract}

\vfill\eject
\section{Introduction}

Hirota bilinear difference equation (HBDE) \cite{Hirota} plays the central role
in the study of integrable nonlinear systems. This single equation embodies
infinite number of integrable differential equations which belong to the
KP-hierarchy as shown by Miwa\cite{Miwa},\cite{Hirota}. It characterizes
algebraic curves\cite{Shiota}. It appears as a consistency relation for the
Laplace maps on the discrete surface\cite{Doliwa}. It is satisfied by string
correlation functions in particle physics\cite{SS}. It is also satisfied by
transfer matrices of some solvable lattice
models\cite{McCoy}\cite{Kuniba}\cite{Korepanov}\cite{KLWZ}.

Solutions to HBDE, called $\tau$ function, are known\cite{Sato}\cite{DJKM} to
form a Grassmann manifold of infinite dimensions. On this space of solutions a
B\"acklund transformation acts and generates the $GL(\infty)$ symmetry. This
large symmetry is the origin of integrability of the system. The scheme of the
B\"acklund transformation can be seen most easily if we convert HBDE into a
pair of linear homogeneous equations. It was first found by Hirota in
\cite{Hirota} and then represented in a manifestly symmetric form under the
B\"acklund transformations in ref.\cite{SSS}. The coupled linear equations
describe the behaviour of a wave function under the influence of gauge
potentials. The compatibility condition for the wave function to solve the
coupled equations requires the gauge potential to satisfy HBDE.

The situation is similar to the inverse scattering method. The difference from
the ordinary inverse method is that every solution to the linear equations also
fulfills the same nonlinear equation satisfied by the gauge potential. This
fact owes to the remarkable symmetry under the exchange of the wave function
and the gauge potential, which we called the dual symmetry in \cite{SSS}.
Hence, by solving the linear system of equations iteratively we obtain a series
of solutions to the nonlinear equation. We call this scheme of finding new
solutions to a nonlinear equation, a linear B\"acklund transformation (LBT).

An interesting application of this scheme was found and described in
\cite{KLWZ}. The authors showed that it can be used to generate Bethe ansatz
solutions to certain class of solvable lattice models. In order to uncover the
meaning of the correlation between the solvable lattice models and the HBDE, it
is desireble to formulate LBT such that the symmetries of this scheme can be
seen from both sides. From this point of view we like to discuss, in this
paper, a symmetric generalization of LBT.

The HBDE is a highly symmetric equation under the exchange of lattice
variables. The linear equations discussed in \cite{SSS}, however, do not
possess this symmetry. In \S 3 we derive a symmetric version of the LBT
associated with HBDE. We will show in \S 4 that we can generalize this scheme
to lattice space of arbitrary dimension. Corresponding to this generalization
we obtain a large number of nonlinear equations which can be solved by the LBT
method. The HBDE turns out to be a special case of this scheme.

\section{Hirota Bilinear Difference Equation (HBDE)}

Let us begin with writing the Hirota bilinear difference equation
(HBDE)\cite{Hirota} for a function $f\in C^\infty$ of discrete variables
$\lambda,\ \mu,\ \nu$:
\begin{eqnarray}
\alpha f(\lambda+1,\mu,\nu)f(\lambda-1,\mu,\nu)&+&\beta
f(\lambda,\mu+1,\nu)f(\lambda,\mu-1,\nu)\nonumber\\
&+&\gamma f(\lambda,\mu,\nu+1)f(\lambda,\mu,\nu-1)
=0
\lb{HBDE}
\end{eqnarray}
where $\alpha,\ \beta, \gamma$ are arbitrary complex parameters. This is a very
simple and highly symmetric equation. In order to show how the LBT generates
solutions to $(\rf{HBDE})$, we follow the argument of ref.\cite{SS}, but in a
different notation convenient later.

We consider the following coupled linear problems:
\begin{equation}
\nabla_{12}\ g=\omega_{12}\ g,\qquad \nabla_{21}\ g=\omega_{21}\ g,
\lb{fg}
\end{equation}
\begin{eqnarray}
\nabla_{12}&:=&f_2 e^{\partial_{\lambda}-\partial_\nu}f_2^{-1}-c_1 f_1
e^{\partial_\mu-\partial_\nu}f_1^{-1}
\nonumber\\
\nabla_{21}&:=&f_1 e^{-\partial_\lambda-\partial_\nu}f_1^{-1}-c_2 f_2
e^{-\partial_{\mu}-\partial_\nu}f_2^{-1},
\end{eqnarray}
where $f_1,\ f_2$ are functions of $\lambda,\ \mu,\ \nu$ and $\omega_{12},\
\omega_{21},\ c_1,\ c_2$ are constants. The shift operator $e^{\partial_x}$
acts to all functions on the right by changing $x$ to $x+1$.

HBDE $(\rf{HBDE})$ emerges from the compatibility condition of this set of
linear equations
\begin{equation}
[\nabla_{12},\ \nabla_{21}]=0.
\lb{[nabla,nabla]}
\end{equation}
In fact, if we impose the condition
$f_1(\lambda,\mu+1,\nu+1)=f_2(\lambda+1,\mu,\nu+1)=:f(\lambda,\mu,\nu)$ to the
gauge potentials, this operator relation turns to
\begin{eqnarray}
&&e^{-2\partial_\nu}{f(\lambda,\mu-1,\nu+1)f(\lambda-1,\mu,\nu+1)\over
f(\lambda-1,\mu-1,\nu)f(\lambda,\mu,\nu)}\left(e^{-\partial_\mu}-e^{-\partial_\lambda}\right)\nonumber\\
&&
\qquad\times \left\{ {f(\lambda+1,\mu,\nu)f(\lambda-1,\mu,\nu)\over
f(\lambda,\mu,\nu+1)f(\lambda,\mu,\nu-1)}
-c_1c_2
{f(\lambda,\mu+1,\nu)f(\lambda,\mu-1,\nu)\over
f(\lambda,\mu,\nu+1)f(\lambda,\mu,\nu-1)}\right\}=0.
\end{eqnarray}
Since the left hand side is a difference of the quantity in the brace, this
quantity itself must be a constant. By adjusting the parameters as
$c_1c_2=-\beta/\alpha$, and putting the constant $-\gamma/\alpha$ we obtain the
HBDE.

An important observation is that the linear equations $(\rf{fg})$ are symmetric
under the exchange of the role of the gauge potentials $f$ and the wave
function $g$. This owes to the fact that under the same constraints for the
guage fields $f$'s, $(\rf{fg})$ can be also written as
\begin{equation}
\tilde\nabla_{12}\ f=\omega_{12}\ f,\qquad \tilde\nabla_{21}\ f=\omega_{21}\ f,
\lb{gf}
\end{equation}
if we define
$g_1(\lambda,\mu-1,\nu-1)=g_2(\lambda-1,\mu,\nu-1)=g(\lambda,\mu,\nu)$ and
\begin{eqnarray}
\tilde\nabla_{12}&:=&g_2 e^{-\partial_{\lambda}+\partial_\nu}g_2^{-1}-c_1 g_1
e^{-\partial_\mu+\partial_\nu}g_1^{-1}
\nonumber\\
\tilde\nabla_{21}&:=&g_1 e^{\partial_\lambda+\partial_\nu}g_1^{-1}-c_2 g_2
e^{\partial_{\mu}+\partial_\nu}g_2^{-1}.
\end{eqnarray}
It is not difficult to convince ourselves that the linear equations $(\rf{gf})$
for the new wave function $f$ can be solved only if the new gauge potential $g$
satisfies $(\rf{HBDE})$, the same equation satisfied by $f$. This is what we
call dual symmetry in \cite{SSS}.

The linear B\"acklund transformation works as follows: Starting from a
particular solution to HBDE, say $f^{(1)}$, as a gauge potential we solve the
linear problem $(\rf{fg})$. One of its solutions, which we call $g^{(1)}$
satisfies the HBDE because it can be regarded as a potential of the coupled
linear equations $(\rf{gf})$ which are nothing but another expression of
$(\rf{fg})$.

The set of coupled equations $(\rf{gf})$ should have solutions other than
$f^{(1)}$. Let us call it $f^{(2)}$. $f^{(2)}$ must also satisfy HBDE, because
$(\rf{fg})$ has $g^{(1)}$ as a solution, hence it is compatible. Now we can
repeat the same argument starting from $f^{(2)}$ to obtain new solution
$g^{(2)}$ to HBDE, and so on. This is an auto B\"acklund transformation since
$f$'s and $g$'s are solutions to the same equation. By starting from the
simplest solution $f^{(1)}=1$ in the case of $\alpha+\beta+\gamma=0$ in
$(\rf{HBDE})$, for instance, we obtain a series of soliton solutions explicitly
in this method.

\section{Symmetrization of LBT}
\vglue 1cm
\begin{center}\unitlength 1mm\begin{picture}(160,60)
\put(0,0){\epsfxsize=14cm\epsfbox{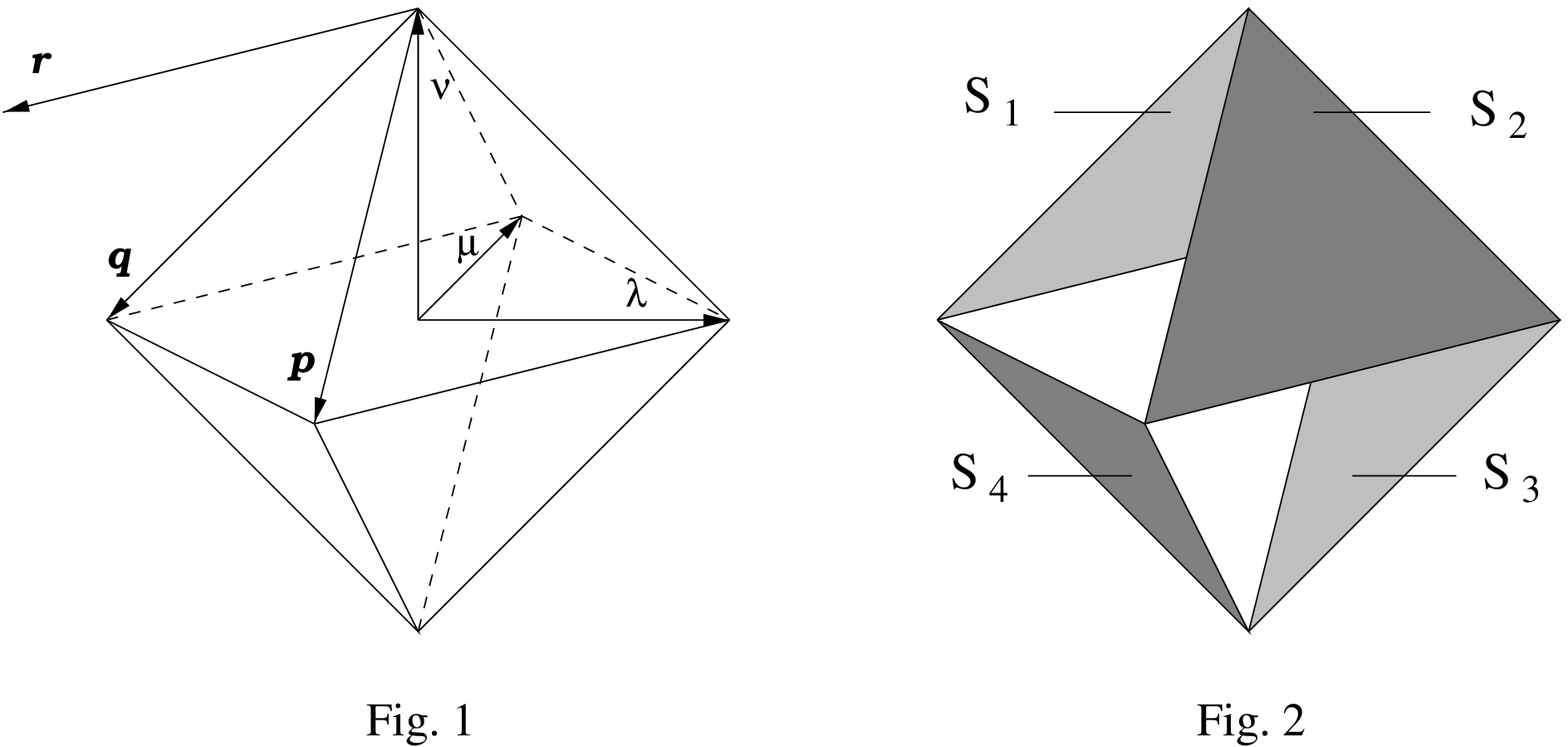}}
\end{picture}\end{center}

The HBDE is highly symmetric by itself, in contrast to its continuous
reductions, such as the KdV equation, Toda lattice equation, sine-Gordon
equation etc.. The corresponding linear version of HBDE, $(\rf{fg})$ and
$(\rf{gf})$, however, are not symmetric. In this section, we like to derive
complete set of linear equations which recover the symmetries possessed by
HBDE.

First we examine the symmetries of the HBDE. From $(\rf{HBDE})$ HBDE relates
functions defined on the six lattice sites $(\lambda\pm 1,\mu,\nu),$
$(\lambda,\mu\pm 1,\nu),$ $ (\lambda,\mu,\nu\pm 1)$, which form an octahedron
in the three dimensional lattice space. Connecting these corners of octahedrons
forms a face centered cubic ({\it fcc}) lattice. Hence dependent variables of
HBDE reside on {\it fcc} lattice, rather than a simple cubic lattice. Indeed
the vectors $\mbox{\boldmath$p$}$,\ $\mbox{\boldmath$q$}$,\
$\mbox{\boldmath$r$}$ in Fig.1 are the primary translation vectors of {\it
fcc}.

The HBDE possesses point group symmetries which transform an octahedron to
itself. It includes the inversion $I$, which changes the sign of the variables
$\lambda,\ \mu,\ \nu$, and the mirror reflections $\sigma_{ij}$ which exchange
the triangles $S_i$ and $S_j$ in Fig. 2. The linear B\"acklund transformation
represented in the form of $(\rf{fg})$, however, is not symmetric under the
group transformation. The purpose of this section is to make it symmetric.

First we notice that the potential field $f_1$ appears in $(\rf{fg})$ as a
gauge field in the form of covariant shift operators
$f_1e^{\partial_\mu-\partial_\nu}f_1^{-1}$ and
$f_1e^{-\partial_\lambda-\partial_\nu}f_1^{-1}$ defined on two edges of the
triangle $S_1$ in Fig. 2. Similarly the field $f_2$ is associated with the
triangle $S_2$. On the other hand from $(\rf{gf})$ we learn that the fields
$g_1$ and $g_2$ are associated with the triangles $\tilde S_1$ and $\tilde
S_2$, respectively, which are obtained from $S_1$ and $S_2$ via inversions with
respect to the origin. Hence two equations $(\rf{fg})$ and $(\rf{gf})$ are
related each other by the inversion $I$. In other words the inversion is
realized by the dual symmetry. The rest of the point group symmetry requires
potential fields associated with $S_3$ and $S_4$ to exist. Let us call them
$f_3$ and $f_4$, respectively.

In order to make the symmetry manifest it is convenient to introduce the
following set of derivatives
\begin{equation}
\hat\partial_1={\partial_\lambda-\partial_\mu-\partial_\nu\over 2},\quad
\hat\partial_2={\partial_\mu-\partial_\nu-\partial_\lambda\over 2},\quad
\hat\partial_3={\partial_\nu-\partial_\lambda-\partial_\mu\over 2},\quad
\hat\partial_4={\partial_\lambda+\partial_\mu+\partial_\nu\over 2}
\end{equation}
They represent the gradients along the normal to the surfaces $S_1,\ S_2,\
S_3,\ S_4$. In terms of these operators the point symmetry transformations can
be expressed as
\begin{eqnarray}
I &:&\quad f_j\leftrightarrow g_j,\quad \hat\partial_j\rightarrow
-\hat\partial_j\nonumber\\
\sigma_{ij} &:&\quad  f_i\leftrightarrow f_j,\quad
\hat\partial_i\leftrightarrow \hat\partial_j.
\lb{symmetry op}
\end{eqnarray}
Applying $\sigma$'s to $(\rf{fg})$ successively we obtain, up to coefficients,
the following six pairs of equations
\begin{eqnarray}
&&
\left(f_je^{\hat\partial_k-\hat\partial_l}f_j^{-1}+f_ke^{\hat\partial_j-\hat\partial_l}f_k^{-1}+1\right)g=0,\nonumber\\
&&
\left(f_ke^{\hat\partial_j-\hat\partial_i}f_k^{-1}+f_je^{\hat\partial_k-\hat\partial_i}f_j^{-1}+1\right)g=0,\nonumber\\
&&
\qquad\qquad j<k,\quad (i,j,k,l)={\rm even\ permutation\ of}\ (1,2,3,4).
\lb{general LBT}
\end{eqnarray}

Among the twelve equations of $(\rf{general LBT})$ we notice that,
corresponding to each triangle of the octahedron, there are three equations
which connect two edges of the same triangle by the gauge fields. For instance
the set of equations $(j,k)=(1,2),\ (2,4),\ (1,4)$ determine the triangle
$S_3$. For the consistency of these equations they must be the same. We can
resolve this problem if we adopt the following conditions to the gauge fields:
\begin{equation}
f_j(\lambda,\mu,\nu)=e^{\hat\partial_j-\hat\partial_4}f(\lambda,\mu,\nu),\qquad
j=1,2,3,4.
\end{equation}
Using the fact that
$\hat\partial_1+\hat\partial_2+\hat\partial_3+\hat\partial_4=0$ we can write
$(\rf{general LBT})$ as
\begin{equation}
\left(e^{-\hat\partial_j-\hat\partial_i}f\right)\left(e^{\hat\partial_j+\hat\partial_4}g\right)+\left(e^{-\hat\partial_k-\hat\partial_i}f\right)\left(e^{\hat\partial_k+\hat\partial_4}g\right)+\left(e^{-\hat\partial_l-\hat\partial_i}f\right)\left(e^{\hat\partial_l+\hat\partial_4}g\right)=0
\end{equation}
for all even permutations of (1,2,3,4). Here we used the notation
$(e^{\hat\partial} f)$ to mean that $e^{\hat\partial}$ acts only to the
functions in the bracket. In this expression it is obvious that, for a given
$i$, three equations with different choices of $j,k,l$ are the same equation.
Therefore we have only four different equations, which we can write as
\begin{equation}
\sum_{j\ne
i}a_{ij}\left(e^{-\hat\partial_i-\hat\partial_j}f\right)\left(e^{\hat\partial_j+\hat\partial_4}g\right)=0,\qquad i=1,2,3,4
\lb{symmetric LBT}
\end{equation}
We have recovered coefficients $a_{ij}$ in this expression. They are free
unless we specify the values of $\alpha,\ \beta,\ \gamma$ in HBDE. This is the
symmetrized LBT which we were looking for and is symmetric under the $\sigma$
transformations.

The symmetric LBT $(\rf{symmetric LBT})$ possesses the $\sigma$ symmetry as we
required. From the analogy of the correspondence between $(\rf{fg})$ and
$(\rf{gf})$, the inversion symmetry, which has not been imposed so far, should
be included if the dual symmetry under the exchange of $f$ and $g$ holds. In
fact if we multiply the shift operator $e^{\hat\partial_i-\hat\partial_4}$ from
the left to the $i$th equation of $(\rf{symmetric LBT})$, it becomes
\begin{equation}
\sum_{j\ne
i}a_{ij}\left(e^{\hat\partial_i+\hat\partial_j}g\right)\left(e^{-\hat\partial_j-\hat\partial_4}f\right)=0,\qquad i=1,2,3,4.
\lb{dual symmetric LBT}
\end{equation}
These are exactly the equations we obtain from $(\rf{symmetric LBT})$ by the
inversion transformation $I$ of $(\rf{symmetry op})$. Therefore $(\rf{symmetric
LBT})$ itself is the totally symmetric LBT which we expected.

Now we will show how HBDE arises from this set of equations. Writing
$(\rf{symmetric LBT})$ explicitely we have, up to the coefficients,
\begin{eqnarray}
\left(\matrix{0&f(\lambda,\mu,\nu+1)&f(\lambda,\mu+1,\nu)&f(\lambda-1,\mu,\nu)\cr
f(\lambda,\mu,\nu+1)&0&f(\lambda+1,\mu,\nu)&f(\lambda,\mu-1,\nu)\cr
f(\lambda,\mu+1,\nu)&f(\lambda+1,\mu,\nu)&0&f(\lambda,\mu,\nu-1)\cr
f(\lambda-1,\mu,\nu)&f(\lambda,\mu-1,\nu)&f(\lambda,\mu,\nu-1)&0\cr}\right)
\left(\matrix{g(\lambda+1,\mu,\nu)\cr g(\lambda,\mu+1,\nu)\cr
g(\lambda,\mu,\nu+1)\cr g(\lambda+1,\mu+1,\nu+1)\cr}\right)&&\nonumber\\
=0.\qquad\qquad&&
\lb{matrix(f)g=0}
\end{eqnarray}
\noindent
Since this is a homogeneous linear equation the determinant of the coefficient
matrix must vanish. For simplicity we assume that the matrix is anti-symmetric,
that is, the coefficients satisfy $a_{ij}=-a_{ji}$. In this case the
determinant is the square of Pfaffian, hence the solvability condition turns
out to be
\begin{eqnarray}
&&
a_{14}a_{23}f(\lambda+1,\mu,\nu)f(\lambda-1,\mu,\nu)-a_{13}a_{24}f(\lambda,\mu+1,\nu)f(\lambda,\mu-1,\nu)\nonumber\\
&&
\qquad\qquad\qquad\qquad
+a_{12}a_{34}f(\lambda,\mu,\nu+1)f(\lambda,\mu,\nu-1)=0.
\end{eqnarray}
In this way the HBDE is reproduced.

\section{Generalization of HBDE to Higher Dimensions}

Our symmetric LBT $(\rf{symmetric LBT})$ possesses the symmetries of the three
dimensional {\it fcc} lattice space. The fourth derivative $\partial_4$,
however, plays a different role in $(\rf{symmetric LBT})$ from others. This
owes to the fact that there are only three independent variables needed to
characterize the four surfaces $S_1\ \sim\ S_4$. This also prevents us to
extend the space to higher dimensions.

We like to show, in this section, that the symmetric LBT $(\rf{symmetric LBT})$
can be rewritten in a manifestly symmetric form in the four dimensional lattice
space if we introduce a parameter $s$ which specifies the order of B\"acklund
transformations. Moreover it also enables us to generalize the linear equations
to arbitrary dimensional lattice space.

Instead of considering two different kinds of fields we associate the parameter
$s+2$ to $f$ and $s+1$ to $g$ and define a new function $f(\lambda,\mu,\nu; s)$
by
\begin{equation}
f(\lambda,\mu,\nu ; s+2)=f(\lambda,\mu,\nu),\qquad f(\lambda,\mu,\nu ;
s+1)=g(\lambda,\mu,\nu).
\end{equation}
We also introduce new variables $k_j$'s with $j=1,2,3,4$ by
$$
k_1:={\lambda-\mu-\nu\over 2}+{s\over 4},\quad k_2:={\mu-\nu-\lambda\over
2}+{s\over 4},\quad k_3:={\nu-\lambda-\mu\over 2}+{s\over 4},
$$
\begin{equation}
 k_4:={\lambda+\mu+\nu\over 2}+{s\over 4}.
\lb{k's}
\end{equation}
It is not difficult to convince ourselves that the operations of
$\hat\partial_j$ is equivalent to the operation of $\partial_j-\partial_s$,
where $\partial_j$ means $\partial/\partial k_j$ for all $j$. Hence the
symmetric LBT $(\rf{symmetric LBT})$ can be written as
\begin{equation}
\sum_{j=1}^4a_{ij}\left(e^{\sum_{l=1}^4\partial_l-\partial_i-\partial_j}f\right)\left(e^{\partial_j}f\right)=0,\qquad i=1,2,3,4,
\lb{symmetric LBT'}
\end{equation}
We notice that $(\rf{symmetric LBT'})$ is totally symmetric for all variables
$k_j,\ j=1,2,3,4$.

In terms of the new variables the HBDE turns out to be
\begin{eqnarray}
&&
\alpha f(k_1+1,k_2,k_3,k_4+1)f(k_1,k_2+1,k_3+1,k_4)\nonumber\\
&&
\qquad\qquad+\beta f(k_1,k_2+1,k_3,k_4+1)f(k_1+1,k_2,k_3+1,k_4)\nonumber\\
&&
\qquad\qquad\qquad\qquad +\gamma
f(k_1,k_2,k_3+1,k_4+1)f(k_1+1,k_2+1,k_3,k_4)=0.
\end{eqnarray}

The introduction of the fourth variable causes nothing to the contents of the
equations. It, however, enables us to generalize equations to higher
dimensional lattice space. We like to show in the rest of this paper that the
generalized equations also provide B\"acklund transformation scheme which
generates solutions to new nonlinear difference equations.

For the convenience we define
\begin{equation}
f_j=e^{\partial_j}f,\qquad \tilde f_j=e^{-\partial_j}f,
\end{equation}
\begin{equation}
\quad
G_{ij}(f)=a_{ij}\exp\left[\sum_{l=1}^n\partial_l-\partial_i-\partial_j\right]f,\qquad \tilde G_{ij}(f)=a_{ij}\exp\left[-\sum_{l=1}^n\partial_l+\partial_i+\partial_j\right]f.
\end{equation}
Then the $n$ dimensional LBT and its dual are
\begin{eqnarray}
&&
\sum_{j=1}^nG_{ij}(f)f_j=0,\qquad i=1,2,3,\cdots,n
\lb{Gg=0}
\\
&&
\sum_{j=1}^n\tilde G_{ij}(f)\tilde f_j=0,\qquad i=1,2,3,\cdots,n.
\lb{Gf=0}
\end{eqnarray}
They are the same equation but written differently. The B\"acklund
transformation of the system $(\rf{Gg=0})$ and $(\rf{Gf=0})$ proceeds in a way
similar to the LBT of  HBDE, which was discussed in \S 2. For the wave function
$f_l$ to be a solution of the homogeneous equation $(\rf{Gg=0})$, the potential
$f$ must satisfy
\begin{equation}
\det\left[G(f)\right]=0.
\lb{det G(f)=0}
\end{equation}
This is a nonlinear difference equation of $f$. At the same time the
solvability condition of $(\rf{Gf=0})$ requires
\begin{equation}
\det\left[\tilde G(f)\right]=0.
\lb{det G(g)=0}
\end{equation}

$(\rf{det G(f)=0})$ and $(\rf{det G(g)=0})$ are the nonlinear equations to be
solved and generalize $(\rf{[nabla,nabla]})$. Since $(\rf{det G(f)=0})$ and
$(\rf{det G(g)=0})$ are not the same equation in general, the transformation
may not be auto B\"acklund transformation. In the case of $n=4$, these two
equations are identical and coincide with the square of HBDE, as we have seen
above. Hence $(\rf{det G(f)=0})$ and $(\rf{det G(g)=0})$ are generalization of
HBDE to $n$ dimensional lattice. Notice that at every step of the B\"acklund
transformation the value of the parameter $s$ is changed by one. Since $s$
specifies a hyper plane in the $n$ dimensional lattice space, the B\"acklund
transformation propagets fields on one plane to the next.

Before closing this paper let us remark briefly the connection of the present
results with other works. A generalization of HBDE to higher dimmensional
lattice space has been discussed in \cite{OHTI}. In their paper the equation
itself is bilinear and is satisfied by the same $\tau$ functions of HBDE. It is
a natural extention because the space of solutions to the KP-hierarchy itself
is symmetric under the choice of three variables out of infinite number of
variables $k_1, k_2, k_3, \cdots$. We have checked that all soliton type of
solutions to the higher dimensional HBDE also satisfy our symmetric LBT.
Moreover we have proved that there exists a solution which satisfies our
trilinear equation but does not satisfy any of bilinear Hirota equations
extended to the four dimensional lattice.

An integrable trilinear partial differential equation (PDE) was discussed in
\cite{MSS}, which has a connection with the Broer-Kaup system. Generalization
to multilinear PDE was also studied in \cite{GRH} and the Painlev\'e properties
were examined. The authors claim that no bilinear form corresponding to them
has been found. Since the equations considered in these references are not
differnce but differential equations, we need further investigation to see the
connection of our work with theirs.


\end{document}